\documentclass[11pt]{article}
\usepackage[margin=1in]{geometry}
\usepackage[utf8]{inputenc}
\usepackage[english]{babel}
\usepackage[T1]{fontenc}
\usepackage{amsmath}
\usepackage{hyperref}
\usepackage{amsthm ,amssymb}

\usepackage{authblk}

\usepackage{caption}
\usepackage{graphicx}
\usepackage{physics}
\usepackage{tikz}

\newtheorem{theorem}{Theorem}[section]

\newtheorem{definition}{Definition}

\begin{document}

\date{\today}
\title{Entanglement Detection by Approximate Entanglement Witnesses}

\author[1]{Samuel Dai}
\author[1,2]{Ning Bao}
\affil[1]{Department of Physics, Northeastern University, Boston, MA 02115, USA}
\affil[2]{Computational Science Initiative, Brookhaven National Laboratory, Upton, NY 11973, USA}

\maketitle

\begin{abstract}
The problem of determining whether a given quantum state is separable is known to be computationally difficult. We propose an approach to this problem based on approximations of convex polytopes in high dimensions. By showing that a convex polytope constructed from a finite number of hyperplanes approximates the Euclidean ball arbitrarily well in high dimensions, we find evidence that a finite set of approximate entanglement witnesses is potentially sufficient to determine the entanglement of a state with high probability.
\end{abstract}

\section{Introduction}

The field of quantum computing has seen significant progress since its conceptual inception in the 1950s. The development of algorithms for factoring \cite{Shor_1997}, quantum simulation \cite{Lloyd_1996}, and solving linear equations \cite{Harrow_Hassidim_Lloyd_2009} are concrete examples where the computational power of quantum computers can potentially surpass that of classical computers. This \emph{quantum advantage} can be said to emerge from phenomena specific to quantum mechanical systems, in particular superposition and entanglement. 

Quantum entanglement is a property of quantum states with multiple subsystems, where the state of the individual subsystems cannot be described without reference to the other systems. Despite its importance, entanglement remains poorly understood; determining whether a given mixed quantum state is separable or entangled is nontrivial even in the simplest bipartite case. For bipartite systems, a solution is known only for $2 \otimes 2$ (two qubits) and $2 \otimes 3$ dimensional (qubit and qutrit) systems, and no solution is known for any higher dimensions \cite{Horodecki_Horodecki_Horodecki_1996}. In fact, the general problem is NP-hard \cite{Gurvits_2003}.

One strategy to approach the separability problem is through the use of entanglement witnesses, operators that have positive expectation values on all separable states, and negative expectation values on at least one entangled state. A natural question to ask is how many or what classes of entangled states can a given witness detect, or how many entanglement witnesses are necessary to detect the entanglement of any entangled state? Unfortunately, it was shown that there does not exist a singular or even finite set of entanglement witnesses that can diagnose entanglement in any quantum state with perfect accuracy \cite{Skowronek_2016}. In fact, at least an exponential number of entanglement witnesses are necessary to detect even a subset of ``robustly'' entangled states \cite{Szarek_Aubrun_2017}\footnote{There have also been numerical approaches to searching for entanglement witnesses for small numbers of qubits, such as \cite{Vintskevich_Bao_Nomerotski_Stankus_Grigoriev_2023}.}. 

In this paper, we will investigate the problem of entanglement detection using a novel method. Motivated by randomized algorithms that obtain greater efficiency by finding approximate solutions \footnote{For example, this occurs when solving semidefinite programming (SDP) problems \cite{Arora_Hazan_Kale_2005}.}, we consider modified entanglement witnesses that can sometimes incorrectly detect entanglement in separable states. We want to determine if it is theoretically possible to gain efficiency in the number of operators needed to detect entanglement by trading off precision in the sense of allowing a small probability of incorrect classification. Using entanglement witnesses with error may be a useful new method of detecting entanglement if the number of witnesses needed decreases substantially compared to using exact witnesses, while maintaining a success probability that is not too small. 

\section{Entanglement Witnesses}

We introduce some background on quantum entanglement. For a more thorough treatment, see the review by the Horodeckis \cite{Horodecki_Horodecki_Horodecki_Horodecki_2009}. 

\subsection{Background}

A $d-$dimensional quantum pure state is a vector $\ket{\psi}$ in a $d-$dimensional Hilbert space $H$. If we consider a system with $n$ composite subsystems, then the total Hilbert space $H$ is given by $H = \otimes_{i=1}^n H_i$, where $H_i$ is the Hilbert space for each individual subsystem\footnote{Note that this assumes factorizability of the Hilbert space, which is guaranteed for qubit systems, though perhaps not for quantum field theories.}. Because the operation connecting the spaces is the tensor product, and not the Cartesian product, it is in general not possible to express the overall quantum state $\ket{\psi}$ as a product of states in each subsystem, i.e. $\ket{\psi} \neq \ket{\psi_1} \otimes \dots \otimes \ket{\psi_n}$\footnote{States that are describable in this form are called \emph{product states}.}, where $\ket{\psi_i}$ is a state vector corresponding to the state of the $i$-th subsystem. In such a case, we call the state $\ket{\psi}$ entangled.

In practice, due to the presence of noise and errors, a more general form for a quantum state is needed, which is the quantum mixed state. A quantum mixed state is a positive semi-definite operator $\rho$ operating on a Hilbert space $H$ with $\tr\rho = 1$ \cite{Nielsen_Chuang_2010}. We say that a mixed state $\rho$ is separable, a natural generalization of the product state form from above, if it can be written as a convex combination of product states, i.e. 
\begin{align}
    \label{sepdef}
    \rho = \sum_i p_i \rho_1^i \otimes \dots \otimes \rho_n^i.
\end{align}
Note that any separable state can be understood as a partial trace over an associated product state. If a mixed state $\rho$ cannot be written in this form, then it is called entangled \cite{Horodecki_Horodecki_Horodecki_1996}. 

Determining whether or not a given mixed state $\rho$ is separable, even in the bipartite case where $n=2$, is in general quite difficult. Note that this is not the case with pure states, where we can simply find the Schmidt decomposition of the vector (state) in question, and ask if the Schmidt rank is one or not \cite{Nielsen_Chuang_2010}. For mixed states, the separability problem has a solution in the bipartite case only for systems of two qubits or systems of one qubit and one qutrit. The solution, due to the Horodeckis, says that a necessary and sufficient condition for a bipartite state $\rho_{AB}$ in $2 \otimes 2$ (two qubits) or $2\otimes 3$ dimensions  (one qubit and one qutrit) to be separable is if the operator $[I \otimes T_B]\rho_{AB}$ is positive \cite{Horodecki_Horodecki_Horodecki_1996}. Here, the operator $T_B$ is the transposition map operating on the second system $B$. This condition is known as the positive partial transpose (PPT) condition \cite{Horodecki_Horodecki_Horodecki_1996}. Unfortunately, the PPT criterion is not a sufficient condition for separability of quantum states in dimensions other than the ones stated above. In particular, states that possess bound entanglement, a form of entanglement where no singlets can be distilled with local operations and classical communication (LOCC), are entangled states that violate the positive partial transpose criterion \cite{Horodecki_Horodecki_Horodecki_1998}.

One fundamental tool in studying mixed state entanglement are entanglement witnesses. Entanglement witnesses are operators that detect entanglement by considering their expectation values on observables \cite{Terhal_2000}. Specifically, an entanglement witness $W$ is a Hermitian operator satisfying
\begin{align}
    \tr(\rho_\text{sep}W) \geq 0 
\end{align}
for all separable states $\rho_\text{sep}$, and 
\begin{align}
    \tr(\rho_\text{ent} W) < 0
\end{align}
for \emph{at least one} entangled state $\rho_\text{ent}$ \cite{Terhal_2000}. A crucial property of entanglement witnesses is that for any entangled state, there necessarily exists an entanglement witness that detects its entanglement. This is due to the Hahn-Banach theorem \cite{Schaefer_H._1999}, which categorizes an entanglement witness as a separating hyperplane between the convex set of separable states and the entangled state detected by it \cite{Terhal_2000}. 

Indeed, this geometric interpretation of entanglement witnesses as a separating is not merely superficial; given an entanglement witness $W$, one can construct another witness $W'$ that detects at least as many entangled states as $W$ by a procedure of subtracting positive operators \cite{Lewenstein_Kraus_Cirac_Horodecki_2000}. By iterating this process, one can construct an optimal entanglement witness $W_\text{opt}$ detecting the maximum number of entangled states possible \cite{Lewenstein_Kraus_Cirac_Horodecki_2000}. Such an entanglement witness is then represented by a hyperplane tangent to the set of separable states. 

\subsection{Approximate Witnesses}

Having discussed the relevant background information on entanglement witnesses, we now introduce the concept of an approximate entanglement witness. 

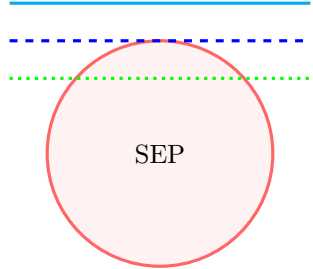
\begin{figure}[ht]
\centering
\begin{tikzpicture}
\filldraw[color=red!60, fill=red!5, very thick](0,0) circle (1.5);
\node (sep) at (0,0) {SEP};
\draw[color=green, very thick, dotted] (-2, 1) -- (2, 1);
\draw[color=blue, very thick, dashed] (-2, 1.5) -- (2, 1.5);
\draw[color=cyan, very thick] (-2, 2.0) -- (2, 2.0);
\end{tikzpicture}
\caption{Geometric illustration of entanglement witnesses and SEP. The solid cyan line indicates a general entanglement witness, the dashed blue line indicates an optimal entanglement witness, and the dotted green line indicates an approximate entanglement witness.}
\end{figure}

\begin{definition}
\label{approx-witness}
An approximate entanglement witness is a Hermitian operator $V$ acting on a Hilbert space $H$ with the following properties:
\begin{enumerate}
    \item There exists a separable state $\rho$ such that $\Tr(\rho  V) \geq 0$.
    \item There exists a separable state $\sigma$ such that $\Tr(\sigma V) < 0 $. 
    \item There exists an entangled state $\tau$ such that $\Tr( \tau V) < 0$. 
\end{enumerate}
\end{definition}

The key difference between an approximate entanglement witness and an exact entanglement witness is that there exist separable states where entanglement is incorrectly witnessed for approximate entanglement witnesses. Geometrically, approximate witnesses are represented by secant hyperplanes that intersect the set of separable states, as opposed to exact witnesses that either do not intersect or are tangent to the set of separable states. 

The existence of an approximate entanglement witness that detects an entangled state follows directly from the existence of an exact entanglement witness that detects said state. An approximate entanglement witness can be constructed from an exact entanglement witness simply by subtracting positive operators, as in \cite{Lewenstein_Kraus_Cirac_Horodecki_2000}. 

\section{The Set of Separable States}

In order for the concept of an approximate entanglement witness to be useful, we need the ability to quantify the accuracy of entanglement detection for a particular approximate witness. Specifically, for a given approximate entanglement witness $W$, we would like to ask how many separable states $\rho$ exist such that $\tr(\rho W) < 0$. Unfortunately, there is no method that currently exists to determine this quantity. While efforts to study the geometry of the set of separable states have been made, such as in calculating its volume  \cite{Zyczkowski_Horodecki_Sanpera_Lewenstein_1998, Szarek_2005} and its inradius \cite{Gurvits_Barnum_2003}, the precise shape of this manifold has not yet been ascertained\footnote{See \cite{Aubrun_Szarek_2017} for a compilation of results in this field}. Hence, we will take an alternative perspective to study this question by transforming the set of separable states to a set that can be studied in greater detail. 

\section{Spherical Transformation}

In this section, we analyze the fraction of separable states that are incorrectly labeled as entangled by considering a toy model of the set of separable states as a unit ball in $\mathbb{R}^d$. 

\begin{figure}[ht]
\centering
\includegraphics[scale=0.4]{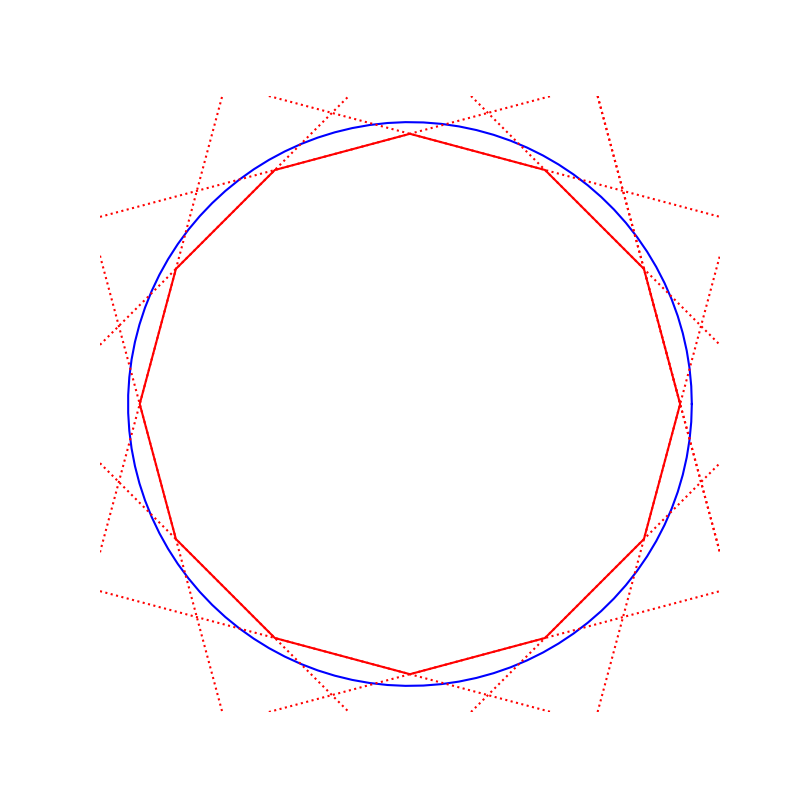}
\caption{Geometric illustration of our approximation strategy in 2 dimensions. The solid blue line represents the boundary of the circle being approximated. The dotted red lines are secant hyperplanes intersecting the circle. The solid red line is the polytope constructed by removing the spherical caps associated with each hyperplane from the circle.}
\end{figure}

The main result is the statement and proof of Theorem \ref{polyapprox}. Since the set of separable states is a compact convex subset of $\mathbb{R}^d$ \cite{Horodecki_1997b}, there exists a homeomorphism that maps it to the closed unit ball $\mathbb{B}^d$. Without finding the explicit transformation, we study the approximate entanglement detection problem by assuming that we have already transformed the set of separable states to a $d$-dimensional ball $\mathbb{B}^d$ in Euclidean space, and that an approximate entanglement witness is a hyperplane in $\mathbb{R}^d$ that intersects $\mathbb{B}^d$. Under this assumption, we show that when the dimension $d$ is large, a finite\footnote{In a previous version of the paper, the authors had mistakenly claimed that Theorem~\ref{polynet} could be used to directly generate a polytope approximation of $\mathbb{B}^d$. This is incorrect because the $\epsilon$-net in Thereom~\ref{polynet} is a \emph{hitting} set, not a \emph{covering} set. Converting to the required covering set requires $\epsilon = O(\exp(d))$.} set of approximate entanglement witnesses is sufficient to approximate the set of separable states\footnote{There is also a potential issue with this approach, as it assumes that the homeomorphism between the ball and the set of separable states takes hyperplanes to hyperplanes, something that is not guaranteed. Nevertheless, we believe that finding entanglement witnesses in this toy model is a useful first step to finding complete sets of approximate witnesses for the set of all separable states.}. 

\begin{theorem}
\label{polyapprox}
There exists a polytope $P^d \subset \mathbb{R}^d$ with $O(\exp(d))$ facets that approximates the unit ball $\mathbb{B}^d \subset \mathbb{R}^d$ so that the ratio between the volume of $P^d$ and the volume of $\mathbb{B}^d$ approaches 1 for sufficiently large $d$. 
\end{theorem}

We comment that the dimension $d$ scaled exponentially with the number of composite subsystems $n$, so that even ``small'' quantum systems exist in a Hilbert space with large dimension. 

The proof of (\ref{polyapprox}) relies on the existence of a polynomial-sized epsilon net for $\mathbb{S}^{n-1}$, which was shown in Theorem 1.2 from \cite{Rabani_Shpilka_2022} by  Rabani and Shpilka. We restate the result below:

\begin{theorem}
\label{polynet}
    There exist two universal constants $a = 2.5$ and $b > 0$ such that for every $\epsilon = \exponential(-O(\sqrt{d}))$ there is an explicit construction of an $\epsilon$-net $S_\epsilon \subset \mathbb{S}^{d-1}$, for spherical caps of size $\abs{S_\epsilon} = O(\epsilon^{-b} \cdot d^a)$.  
\end{theorem}

\subsection{Proof of Theorem \ref{polyapprox}}

\begin{proof}
Let $0 < \epsilon \ll 1$ be fixed. Then by Theorem \ref{polynet}, there exists an epsilon net $S_\epsilon \subset \mathbb{S}^{d-1}$ of size $\abs{S_\epsilon} = O(\epsilon^{-b} \cdot d^a)$. Denote the points in $S_\epsilon$ by $y_i$, for $i \in \{1, \dots, \abs{S_\epsilon}\}$. Consider the hyperplanes defined for each $y_i \in S_\epsilon$ by $H_i = \{ x \in \mathbb{R}^d: \langle x, y_i \rangle = \cos\epsilon \}$. Each hyperplane intersects the sphere to form a spherical cap defined by $C_i = \{ x \in \mathbb{B}^d : \langle x, y_i \rangle < \cos\epsilon\}$. We construct a polytope $P^d$ contained in $\mathbb{B}^d$ using the spherical caps $C_i$ by 
\begin{equation}
    P^d = \mathbb{B}^d \setminus \bigcup_{y_i \in S_\epsilon} C_i.
\end{equation}
A lower bound for the volume of the polytope $P^d$ is then 
\begin{equation}
\label{polytope_expression}
    \text{vol}(P^d) \geq \text{vol}(\mathbb{B}^d) - \sum_{y_i \in S_\epsilon} \text{vol}(C_i).
\end{equation}

The volume of a $d$-dimensional ball with radius $r$, denoted by $B_r^d$, is known to be
\begin{equation}
    \text{vol}(B_r^d) = \frac{\pi^{d/2}}{\Gamma(1 + d/2)}r^d.
\end{equation}

For the unit ball, we have 
\begin{equation}
    \text{vol}(\mathbb{B}^d)= \frac{\pi^{d/2}}{\Gamma(1 + d/2)}.
\end{equation}

The volume of the spherical cap $C_i$ can thus be found by integrating the volume of a $(d-1)$-dimensional ball with radius $\sin\theta$ from $\theta=0$ to $\theta =\epsilon$ \cite{Li_2010}: 
\begin{align}
    \text{vol}(C_i) &= \int_0^\epsilon \text{vol}(B^{d-1}_{\sin\theta}) \sin\theta d\theta \\ 
    &= \frac{\pi^{(d-1)/2}}{\Gamma(1 + (d-1)/2)} \int_0^\epsilon \sin^d \theta d\theta. 
\end{align}
Substituting the expressions for $\text{vol}(\mathbb{B}^d)$ and $\text{vol}(C_i)$ into $(\ref{polytope_expression})$, we see that the ratio between the volume of $\text{vol}(P^d)$ and $\text{vol}(\mathbb{B}^d)$ is bounded below by 
\begin{equation}
\label{vol-ratio}
    \frac{\text{vol}(P^d)}{\text{vol}(\mathbb{B}^d)} \geq 1 - \abs{S_\epsilon}\frac{\Gamma(1+d/2)}{\pi^{1/2} \Gamma(1 + (d-1)/2)} \int_0^\epsilon \sin^d \theta d\theta.
\end{equation}

Using a small angle approximation for $\sin\theta$ and a Stirling-type approximation for the Gamma functions \cite{Gordon_1994}, we can further approximate the volume ratio in (\ref{vol-ratio}). 

We use the following bounds on the Gamma function from \cite{Gordon_1994}: 
\begin{equation}
   \sqrt{2\pi} d^{d-\frac{1}{2}} e^{-d} e^{\frac{1}{12d + 6/7}}< \Gamma(d) < \sqrt{2\pi} d^{d-\frac{1}{2}} e^{-d} e^{\frac{1}{12d}},
\end{equation}
which are valid for $d > 0$. 

Applying these bounds,
\begin{align}
    \Gamma(\frac{d}{2}+1) &< \sqrt{2\pi} \qty(\frac{d}{2}+1)^{\frac{d+1}{2}} e^{-(\frac{d}{2}+1)} e^{\frac{1}{6d+12}},\\
    \Gamma(\frac{d+1}{2}) &>  \sqrt{2\pi} \qty(\frac{d+1}{2})^{\frac{d}{2}} e^{-\frac{d+1}{2}} e^{\frac{1}{6d + 48/7}},
\end{align}
and so 
\begin{align}
    \frac{\Gamma(1 +d/2)}{\Gamma(1 + (d-1)/2)} &< \frac{\qty(\frac{d+2}{2})^{\frac{d+1}{2}} e^{-(\frac{d}{2}+1)} e^{\frac{1}{6d+12}}}{\qty(\frac{d+1}{2})^{\frac{d}{2}} e^{-\frac{d+1}{2}} e^{\frac{1}{6d + 48/7}}} \\ 
    &=  \qty(\frac{d+2}{2})^{1/2}\qty(\frac{d+2}{d+1})^{d/2} e^{-1/2} e^{-\frac{1}{(d+2)(7d+8)}} \\ 
    &= \frac{1}{(2e)^{1/2}}\qty(\frac{d+2}{d+1})^{d/2} e^{-\frac{1}{(d+2)(7d+8)}} (d+2)^{1/2}.
\end{align}
For $0 < \theta \ll 1$, we have $0 < \sin\theta < \theta$. Hence, 
\begin{align}
    \int_0^\epsilon \sin^d \theta d\theta &< \int_0^\epsilon \theta^d d\theta = \frac{\epsilon^{d+1}}{d+1}.
\end{align}
We see that the volume ratio in equation (\ref{vol-ratio}) can be approximated by 
\begin{align}
    \frac{\text{vol}(P^d)}{\text{vol}(\mathbb{B}^d)} > 1 - \abs{S_\epsilon}\frac{1}{(2e\pi)^{1/2}}\qty(\frac{d+2}{d+1})^{d/2} e^{-\frac{1}{(d+2)(7d+8)}} \frac{(d+2)^{1/2}}{d+1} \epsilon^{d+1}.
\end{align}


The result follows upon inserting $\abs{S_\epsilon} = O(\exp(d))$. 
\end{proof}

We comment that this method constructs a polytope approximation contained inside SEP. In terms of entanglement detection, errors in the approximation correspond to states that are incorrectly identified as entangled when they are actually separable. However, the same method with a slight modification constructs a polytope approximation that contains SEP (simply take hyperplanes that intersect outside of the ball instead of on the inside). Here, errors in the approximation would correspond to states that are incorrectly identified as separable when they are actually entangled. Thus, which type of error is acceptable is a matter of preference -- the size of the approximating polytope is the same in either case.  

\section{Discussion}

We have explored the entanglement detection problem through the study of approximate entanglement witnesses. After constructing a toy model of the compact convex set of separable states as a ball $\mathbb{B}^d$, we constructed a convex polytope by removing the spherical caps associated with the intersection of a finite number of hyperplanes from the unit ball $\mathbb{B}^d$.  We then showed that the difference in volume between the constructed polytope and $\mathbb{B}^d$ becomes arbitrarily small as the dimension becomes large. 

However, as transforming the set of separable states to a unit ball will not generically cause entanglement witnesses to remain hyperplanes once transformed, since it is in general not true that a hyperplane in one space will remain a hyperplane in a homeomorphic space, we cannot directly apply our result to the set of separable states. Nevertheless, it may be possible to apply our result using alternative methods. 

We discuss one potential method. First, observe that a ball in $\mathbb{R}^D$ can be transformed into an ellipsoid by an invertible linear map. Moreover, such a transformation maps hyperplanes to hyperplanes, so a simple consequence of Theorem~\ref{polyapprox} is that ellipsoids in $\mathbb{R}^D$ can be approximated using a finite number of hyperplanes. One can then approximate SEP by constructing a union of a finite number of ellipsoids, each of which can be approximated using a finite number of hyperplanes\footnote{Note that extraneous hyperplanes can be discarded on the interior of the set to avoid extra misidentification.}. We would then be able to conclude that SEP can be approximated using a finite number of approximate entanglement witnesses. We leave this for future work.

To conclude, while we did not find the explicit homeomorphism between the set of separable states and the closed unit ball, our result speaks positively about the potential usefulness of approximate entanglement witnesses. We believe that it is likely that only a finite number of approximate witnesses is necessary to detect entangled states with high probability, compared to the super-exponential number of exact entanglement witnesses necessary. In this context, it would be useful to study the geometry of the set of separable states for further investigation. We also emphasize that our results are centered around the theoretical advantages that may exist in using approximate entanglement witnesses. We leave development of practical methods to future work. 

\section*{Acknowledgements}
We thank Nathaniel Johnston, Debbie Leung, and Benjamin Lovitz for useful discussions on this topic. N.B. is supported by the Computational Science Initiative at Brookhaven National Laboratory, Northeastern University, and by the U.S. Department of Energy QuantISED Quantum Telescope award.

\bibliographystyle{alpha}
\bibliography{ref}

@article{Li_2010, title={Concise formulas for the area and volume of a hyperspherical cap}, volume={4}, DOI={10.3923/ajms.2011.66.70}, number={1}, journal={Asian Journal of Mathematics \& Statistics}, author={Li, S.}, year={2010}, pages={66–70}}

@article{Gordon_1994, title={A stochastic approach to the gamma function}, volume={101}, DOI={10.2307/2975134}, number={9}, journal={The American Mathematical Monthly}, author={Gordon, Louis}, year={1994}, pages={858}}

@article{Szarek_Aubrun_2017, title={Dvoretzky’s theorem and the complexity of entanglement detection}, DOI={10.19086/da.1242}, journal={Discrete Analysis}, author={Szarek, Stanislaw and Aubrun, Guillaume}, year={2017}, pages={1–20}}

@article{Horodecki_Horodecki_Horodecki_1996, title={Separability of mixed states: Necessary and sufficient conditions}, volume={223}, DOI={10.1016/s0375-9601(96)00706-2}, number={1–2}, journal={Physics Letters A}, author={Horodecki, Michał and Horodecki, Paweł and Horodecki, Ryszard}, year={1996}, pages={1–8}}

@article{Skowronek_2016, title={There is no direct generalization of positive partial transpose criterion to the three-by-three case}, volume={57}, DOI={10.1063/1.4966984}, number={11}, journal={Journal of Mathematical Physics}, author={Skowronek, Łukasz}, year={2016}}

@article{Lewenstein_Kraus_Cirac_Horodecki_2000, title={Optimization of entanglement witnesses}, volume={62}, DOI={10.1103/physreva.62.052310}, number={5}, journal={Physical Review A}, author={Lewenstein, M. and Kraus, B. and Cirac, J. I. and Horodecki, P.}, year={2000}}

@article{Gurvits_2003, title={Classical deterministic complexity of Edmonds’ problem and quantum entanglement}, DOI={10.1145/780542.780545}, journal={Proceedings of the thirty-fifth annual ACM symposium on Theory of computing}, author={Gurvits, Leonid}, year={2003}}

@article{Rabani_Shpilka_2022, title={Corrigendum: Explicit construction of a small epsilon-net for linear threshold functions}, volume={51}, DOI={10.1137/20m1310321}, number={5}, journal={SIAM Journal on Computing}, author={Rabani, Yuval and Shpilka, Amir}, year={2022}, pages={1692–1702}}

@article{Harrow_Hassidim_Lloyd_2009, title={Quantum algorithm for linear systems of equations}, volume={103}, DOI={10.1103/physrevlett.103.150502}, number={15}, journal={Physical Review Letters}, author={Harrow, Aram W. and Hassidim, Avinatan and Lloyd, Seth}, year={2009}, month={Oct}}

@article{Shor_1997, title={Polynomial-time algorithms for prime factorization and discrete logarithms on a quantum computer}, volume={26}, DOI={10.1137/s0097539795293172}, number={5}, journal={SIAM Journal on Computing}, author={Shor, Peter W.}, year={1997}, month={Oct}, pages={1484–1509}}

@article{Lloyd_1996, title={Universal quantum simulators}, volume={273}, DOI={10.1126/science.273.5278.1073}, number={5278}, journal={Science}, author={Lloyd, Seth}, year={1996}, month={Aug}, pages={1073–1078}}

@article{Terhal_2000, title={Bell inequalities and the separability criterion}, volume={271}, DOI={10.1016/s0375-9601(00)00401-1}, number={5–6}, journal={Physics Letters A}, author={Terhal, Barbara M.}, year={2000}, month={Jul}, pages={319–326}}

@book{Nielsen_Chuang_2010, place={Cambridge}, title={Quantum Computation and Quantum Information: 10th Anniversary Edition}, publisher={Cambridge University Press}, author={Nielsen, Michael A. and Chuang, Isaac L.}, year={2010}}

@article{Horodecki_Horodecki_Horodecki_1998, title={Mixed-state entanglement and distillation: Is there a “bound” entanglement in nature?}, volume={80}, DOI={10.1103/physrevlett.80.5239}, number={24}, journal={Physical Review Letters}, author={Horodecki, Michał and Horodecki, Paweł and Horodecki, Ryszard}, year={1998}, month={Jun}, pages={5239–5242}}

@article{Horodecki_1997b, title={Separability criterion and inseparable mixed states with positive partial transposition}, volume={232}, DOI={10.1016/s0375-9601(97)00416-7}, number={5}, journal={Physics Letters A}, author={Horodecki, Pawel}, year={1997}, month={Aug}, pages={333–339}}

@article{Zyczkowski_Horodecki_Sanpera_Lewenstein_1998, title={Volume of the set of separable states}, volume={58}, DOI={10.1103/physreva.58.883}, number={2}, journal={Physical Review A}, author={{\.Z}yczkowski, Karol and Horodecki, Paweł and Sanpera, Anna and Lewenstein, Maciej}, year={1998}, month={Aug}, pages={883–892}}

@article{Szarek_2005, title={Volume of separable states is super-doubly-exponentially small in the number of qubits}, volume={72}, DOI={10.1103/physreva.72.032304}, number={3}, journal={Physical Review A}, author={Szarek, Stanislaw J.}, year={2005}, month={Sep}}

@article{Gurvits_Barnum_2003, title={Separable balls around the maximally mixed multipartite quantum states}, volume={68}, DOI={10.1103/physreva.68.042312}, number={4}, journal={Physical Review A}, author={Gurvits, Leonid and Barnum, Howard}, year={2003}, month={Oct}}

@article{Vintskevich_Bao_Nomerotski_Stankus_Grigoriev_2023, title={Classification of four-qubit entangled states via machine learning}, volume={107}, DOI={10.1103/physreva.107.032421}, number={3}, journal={Physical Review A}, author={Vintskevich, S. V. and Bao, N. and Nomerotski, A. and Stankus, P. and Grigoriev, D. A.}, year={2023}, month={Mar}}

@book{Schaefer_H._1999, place={New York}, title={Topological vector spaces}, publisher={Springer}, author={Schaefer, Helmut H. and H., Wolff Manfred P}, year={1999}}

@book{Aubrun_Szarek_2017, place={Providence (R.I.)}, title={Alice and Bob Meet Banach: The interface of asymptotic geometric analysis and Quantum Information theory}, publisher={American Mathematical Society}, author={Aubrun, Guillaume and Szarek, Stanislaw}, year={2017}}

@article{Arora_Hazan_Kale_2005, title={Fast algorithms for approximate semidefinite programming using the multiplicative weights update method}, DOI={10.1109/sfcs.2005.35}, journal={46th Annual IEEE Symposium on Foundations of Computer Science (FOCS’05)}, author={Arora, S. and Hazan, E. and Kale, S.}, year={2005}}

@article{Horodecki_Horodecki_Horodecki_Horodecki_2009, title={Quantum Entanglement}, volume={81}, DOI={10.1103/revmodphys.81.865}, number={2}, journal={Reviews of Modern Physics}, author={Horodecki, Ryszard and Horodecki, Paweł and Horodecki, Michał and Horodecki, Karol}, year={2009}, month={Jun}, pages={865–942}}

\end{document}